\documentclass[aip,pop,reprint,showpacs,twocolumn,amsmath,amssymb]{revtex4-1}
\usepackage[]{graphicx}
\usepackage{amsmath}
\usepackage{color}
\usepackage{enumerate}

\begin{document}

\title{Wave mode coupling due to plasma wakes in two-dimensional plasma crystals: In-depth view}

\author{L. Cou\"edel}
\affiliation{Max Planck Institute for Extraterrestrial Physics, 85741 Garching, Germany}
\affiliation{PIIM, CNRS/Universit\'e de Provence, 13397 Marseille Cedex 20, France}
\author{S. K. Zhdanov}
\affiliation{Max Planck Institute for Extraterrestrial Physics, 85741 Garching, Germany}
\author{A. V. Ivlev}\email{ivlev@mpe.mpg.de}
\affiliation{Max Planck Institute for Extraterrestrial Physics, 85741 Garching, Germany}
\author{V. Nosenko}
\affiliation{Max Planck Institute for Extraterrestrial Physics, 85741 Garching, Germany}
\author{H. M. Thomas}
\affiliation{Max Planck Institute for Extraterrestrial Physics, 85741 Garching, Germany}
\author{G. E. Morfill}
\affiliation{Max Planck Institute for Extraterrestrial Physics, 85741 Garching, Germany}

\begin{abstract}
Experiments with two-dimensional (2D) plasma crystals are usually carried out in rf plasma sheaths, where the interparticle
interactions are modified due to the presence of plasma wakes. The wake-mediated interactions result in the coupling between
wave modes in 2D crystals, which can trigger the mode-coupling instability and cause melting. The theory predicts a number
of distinct fingerprints to be observed upon the instability onset, such as the emergence of a new hybrid mode, a critical
angular dependence, a mixed polarization, and distinct thresholds. In this paper we summarize these key features and provide
their detailed discussion, analyze the critical dependence on experimental parameters, and highlight the outstanding issues.
\end{abstract}


\maketitle

\section{Introduction}

Stability of two-dimensional (2D) plasma crystals is a fundamental problem of complex (dusty) plasmas. Such crystals --
which are monolayers of hexagonally ordered monodisperse microparticles -- can be (routinely) created in a rf plasma
\cite{Fortov20051,RevModPhys.81.1353}: Particles get negatively charged in a plasma and therefore the electric force exerted
on them in the sheath above a horizontal electrode can compensate for gravity, thus providing a stable levitation. There are
several mechanisms operating in complex plasmas that can result in the melting of 2D crystals. These mechanisms can
generally be divided into two categories -- {\it generic} and {\it plasma-specific}.

Generic mechanisms of melting are those operating in any (classical) system with a given pair interaction between particles,
provided the interaction can be described by a Hamiltonian. The melting in 2D systems can either be a two-step process
(which involves consecutive unbinding of weakly interacting dislocation and disclination pairs, respectively, with the
intermediate hexatic phase) \cite{PhysRevLett.41.121,0022-3719-6-7-010,PhysRevB.19.2457,PhysRevB.19.1855} or a one-step
process (where the hexatic phase is preempted by the formation of dislocation chains) \cite{PhysRevB.28.178}. These generic
melting mechanisms can operate in very different 2D systems including complex plasmas
\cite{PhysRevLett.58.1200,nosenko:nzikm,PhysRevE.53.R2049,sheridan:083705,sheridan:103702,PhysRevLett.85.3656}.

Plasma-specific melting mechanisms can only operate in complex plasmas. Such mechanisms are associated with the energy
exchange between microparticles and ambient plasma and can be considered as a result of the system openness. For instance,
2D plasma crystals can be strongly perturbed by single particles moving above or below the monolayer
\cite{PhysRevE.62.1238,Nunomura05}, or they can melt due to fluctuations of particle charges
\cite{PhysRevLett.83.1970,Morfill99,vaulina:vknp}.

The most universal among the plasma-specific mechanisms is that associated with the {\it wake-mediated} interaction between
microparticles: In the presence of strong plasma flow the screening cloud around each charged particle becomes highly
asymmetric \cite{ishihara:357,ishihara:69,vladimirov:3867,vladimirov:vs}. These clouds are usually referred to as ``plasma
wakes'' \cite{Lampe00,Miloch08,PhysRevE.64.046406,kompaneets:052108,Kompaneets08} and play the role of a ``third body'' in
the interparticle interaction, making it nonreciprocal \cite{PhysRevLett.83.3194}. Under certain conditions, this makes the
system non-Hamiltonian and provides effective conversion of the energy of flowing ions into the kinetic energy of
microparticles \cite{Fortov20051,RevModPhys.81.1353,PhysRevE.54.R46,Joyce02}.

The wake-induced mechanism of melting of crystalline monolayers was discovered theoretically a decade ago by Ivlev and
Morfill \cite{PhysRevE.63.016409}. Based on a simple model of a particle chain, it was shown that the longitudinal in-plane
and transverse out-of-plane dust-lattice (DL) wave modes are no longer independent -- they are coupled due to the
wake-mediated interactions. When the modes intersect they become modified by the coupling and form a {\it hybrid} mode in a
narrow vicinity of the crossing. This can trigger the {\it mode-coupling instability} which causes the melting. This melting
mechanism had received strong confirmation later on, when the instability threshold predicted by the theory was compared
\cite{PhysRevE.68.026405} with the experimental observations by U. Konopka and numerical simulations by G. Joyce. However,
the direct comparison of theory and experiment became possible only recently, after an experimental method of measuring the
out-of-plane mode was developed \cite{couedel:215001} and therefore the essential fingerprint of the mode-coupling
instability -- the hybrid mode -- became observable. In recent experiment by Cou\"edel {\it et al.}
\cite{PhysRevLett.104.195001} an implementation of this method unambiguously demonstrated that the melting indeed occurs due
to the resonance coupling between the longitudinal in-plane and transverse out-of-plane modes. The variation of the wave
modes with the experimental conditions, including the emergence of the hybrid branch, revealed exceptionally good agreement
with the theory of mode-coupling instability generalized for 2D case by Zhdanov {\it et al.} \cite{zhdanov:zim}.

The theory of mode-coupling instability \cite{PhysRevE.68.026405,PhysRevE.63.016409,zhdanov:zim} predicts a number of
distinct fingerprints to be observed upon the instability onset: Along with the emergence of hybrid mode mentioned above,
these are a critical angular dependence -- the hybrid mode first appears only for wave vectors oriented along one of the
principal lattice axes, a mixed polarization -- the two DL modes that form the hybrid mode are no longer purely longitudinal
and transverse close to the merging point, and distinct thresholds -- the instability sets in only when (i) the particle
number density in the monolayer is high enough or/and vertical confinement eigenfrequency is low enough (so that the two DL
modes can cross and form the hybrid mode) and (ii) the gas pressure is low enough (so that the growth rate of the hybrid
mode exceeds the damping rate).

All these fingerprints have been mentioned in our previous theoretical publications
\cite{PhysRevE.63.016409,PhysRevE.68.026405,zhdanov:zim}, some of them were also illustrated in the followup experimental
paper \cite{PhysRevLett.104.195001}. However, their discussion was apparently too concise to address all these important
properties in necessary detail. The need for in-depth discussion of the wake-mediated mode coupling became evident now, when
new publications have appeared where some essential properties of the instability were misinterpreted (see, e.g., recent
experimental paper \cite{PhysRevLett.105.085004} and the subsequent Erratum \cite{PhysRevLett.105.269901} by Liu {\it et
al.}). Therefore, in this paper we summarize the key features of the mode coupling and provide a detailed discussion,
analyze the critical dependence on experimental parameters, and highlight the outstanding issues.

\section{Summary of Mode-Coupling Theory}\label{theory}

In 2D plasma crystals, two in-plane wave modes can be sustained (here we naturally leave aside polydisperse mixtures). Both
modes have an acoustic dispersion, one of them is longitudinal and the other is transverse. Since the strength of the
vertical confinement in such systems is finite, there is a third fundamental wave mode which has an optical dispersion and
is associated with the out-of-plane oscillations
\cite{PhysRevLett.71.2753,PhysRevE.63.016409,PhysRevE.68.046403,samsonov:026410,PhysRevE.56.R74,vladimirov:030703,zhdanov:zim}.
Theory predicts that all three modes depend critically on the parameters of the plasma wakes. Yet the role of the wakes is
not only the modification of the modes themselves. What is far more important is that the modes are {\it coupled} to each
other due to the wake-mediated interactions between particles. It is worth noting that this {\it coupling is linear} and
therefore does not depend on the wave amplitude. Below we identify the conditions when the wake-mediated coupling becomes
crucial for waves in 2D crystals.

\subsection{DL modes for nonreciprocal interactions}

In general, linear dispersion relations $\omega({\bf k})$ are determined by eigenvalues (eigenmodes) of a dynamical matrix
$\textsf{D}$. The latter is derived by considering small perturbations of individual particles (with respect to a stable
configuration, e.g., a hexagonal monolayer for 2D crystals) of the form $\propto\exp(-i\omega t+i{\bf k}\cdot{\bf r})$.
Elements of $\textsf{D}$ are determined by properties of interparticle interactions and are functions of ${\bf k}$. If the
interactions are reciprocal then $\textsf{D}$ is Hermitian, and therefore the eigenvalues are always real (i.e., the modes
are stable) and the eigenvectors are orthogonal. The situation changes if the interactions are nonreciprocal -- the
eigenvalues in this case can become complex.

\begin{figure}
\includegraphics[width=8.3cm,clip=]{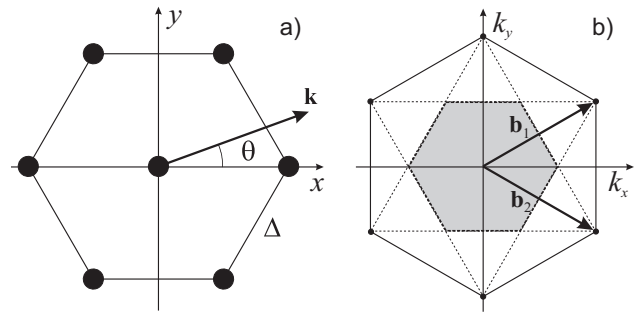}
\caption{\label{fig1} (a) Elementary hexagonal lattice cell with the frame of reference, the lattice constant is $\Delta$.
(b) The reciprocal lattice in ${\bf k}$-space, the basis vectors of the lattice are ${\bf b}_{1,2}=2\pi\Delta^{-1} (1,\pm
\frac1{\sqrt{3}})$. Due to the lattice symmetry, it is sufficient to consider the wave vectors ${\bf k}$ at $0^{\circ}\leq
\theta\leq30^{\circ}$ and from within the first Brillouin zone (gray region enclosed by dashed lines), so that $|{\bf k}|
\Delta\leq\frac43\pi$ for $\theta=0^{\circ}$ and $|{\bf k}|\Delta\leq\frac2{\sqrt{3}}\pi$ for $\theta=30^{\circ}$.}
\end{figure}

Before specifying a particular functional form of interparticle interactions in 2D crystals, let us generally assume the
non-reciprocity and discuss common properties of the DL waves in this case. The dynamical matrix has the following form:
\begin{equation}\label{matrix}
\textsf{D}=
\left(
  \begin{array}{ccc}
    \alpha_{\rm h}-\beta & 2\gamma & i\sigma_y \\
    2\gamma & \alpha_{\rm h}+\beta & i\sigma_x \\
    i\sigma_y & i\sigma_x & \Omega_{\rm conf}^2-2\alpha_{\rm v} \\
  \end{array}
\right).
\end{equation}
The elements $\alpha_{\rm h}({\bf k})$, $\beta({\bf k})$, and $\gamma({\bf k})$ determine the dispersion of two in-plane
(horizontal) modes and $\alpha_{\rm v}({\bf k})$ characterizes the out-of-plane (vertical) mode. The elements
$\sigma_{x,y}({\bf k})$ emerge due to non-reciprocity and make $\textsf{D}$ non-Hermitian. The matrix is calculated for the
reference frame shown in Fig.~\ref{fig1}(a) assuming that vertically the particles are confined in a (parabolic) potential
well characterized by the eigenfrequency $\Omega_{\rm conf}=2\pi f_{\rm conf}$. To derive the dispersion relations
$\omega({\bf k})$, we add a friction force $m\nu\dot{\bf r}$ in equations of motion (with $\nu$ being the damping rate)
which yields:
\begin{equation}\label{dispersion}
    \det[\textsf{D} -\omega(\omega+i\nu)\textsf{I}]=0,
\end{equation}
where $\textsf{I}$ is the unit matrix. Thus, $\omega(\omega+i\nu)\equiv\Omega^2$ are the eigenvalues of $\textsf{D}$, i.e.,
the DL wave modes.

For reciprocal interactions $\sigma_{x,y}=0$ and then Eqs (\ref{matrix}) and (\ref{dispersion}) yield
\begin{equation}\label{dispersion1}
    (\Omega^2-\Omega_{{\rm h}\|}^2)(\Omega^2-\Omega_{{\rm h}\perp}^2)(\Omega^2-\Omega_{\rm v}^2)=0.
\end{equation}
There are three independent and real DL modes: Two {\it acoustic} in-plane modes $\Omega^2({\bf k})=\alpha_{\rm
h}\pm\sqrt{\beta^2+4\gamma^2}\equiv\Omega_{{\rm h}\|,\perp}^2({\bf k})$ (where $\|$ and $\perp$ indicate the longitudinal
and transverse polarizations and correspond to the plus and minus signs, respectively) and an {\it optical} out-of-plane
mode $\Omega^2({\bf k})=\Omega_{\rm conf}^2-2\alpha_{\rm v}\equiv\Omega_{\rm v}^2({\bf k})$. When solved for $\omega({\bf
k})$, each mode yields a couple of conjugate branches.

When interactions are nonreciprocal the modes are modified -- they are described by Eq. (\ref{dispersion1}) with nonzero
right-hand side (r.h.s.) which is proportional to $\sigma_x^2+\sigma_y^2$. The DL modes become coupled with each other and
complex \footnote{Matrix $\textsf{D}$ is symmetric and hence can be reduced to the diagonal form also for
$\sigma_{x,y}\ne0$, but with complex eigenvalues.}.

Now let us specify interparticle interactions. For the direct interaction between charged grains we naturally adopt the
Yukawa potential which is characterized by (negative) particle charge $Q$ and the screening length $\lambda$. For the wake
we employ the commonly used simple model \cite{PhysRevE.54.R46} of a point-like (positive) charge $q$ located at the
distance $\delta$ downstream each particle and also interacting with the neighboring particles via the Yukawa potential. The
elements of the dynamical matrix (\ref{matrix}) for such interactions are calculated in Appendix~\ref{appendix1},
Eqs~(\ref{elements1}) and (\ref{elements2}). For brevity, theoretical results in this paper are written in the dimensionless
form -- the frequency is normalized by the DL frequency scale,
\begin{displaymath}
    \Omega_{\rm DL}=\sqrt{\frac{Q^2}{m\lambda^3}},
\end{displaymath}
and the wave vector is normalized by the lattice constant $\Delta$: $\Omega/\Omega_{\rm DL}\to\Omega$ and ${\bf
k}\Delta\to{\bf k}$.

\subsection{Types of mode coupling}\label{types}

After the normalization, the DL modes depend on three dimensionless parameters (along with the normalized wave vector): The
screening parameter $\kappa=\Delta/\lambda$ as well as the relative wake charge $\tilde q=|q/Q|$ and distance
$\tilde\delta=\delta/\Delta$. In experiments, $\kappa$ is typically about unity whereas both wake parameters are usually
small. Elements $\sigma_{x,y}$ [Eq.~(\ref{elements2})] are proportional to the product $\tilde q\tilde\delta$ and hence the
coupling between the DL modes is very small [$\propto(\tilde q\tilde\delta)^2$]. It becomes important only if different DL
modes intersect (or they are very close to each other); otherwise the coupling can be neglected and then the modes can be
treated independently, as described by Eq. (\ref{dispersion1}). Let us elaborate on this point.

\begin{figure}
\includegraphics[width=0.99\columnwidth,clip=]{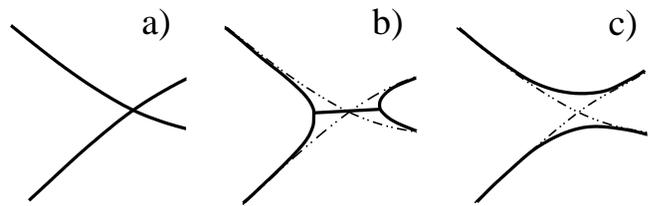}
\caption{\label{fig2} Modification of coupled modes [Eq. (\ref{dispersion2})] in the vicinity of their crossing. The sketch
depicts the modes in the $(\Omega,|{\bf k}|)$-plane (a) in the absence of coupling, (b) with negative coupling ($\epsilon<0$),
and (c) with positive coupling ($\epsilon>0$). Negative coupling results in the formation of a hybrid mode which can be unstable
(see text); positive coupling causes the mode reconnection, while the modes remain stable and form a ``forbidden band''.}
\end{figure}

Behavior of two coupled modes, $\Omega_{1,2}({\bf k})$, in a close proximity of their intersection can generally be
described by the following equation:
\begin{equation}\label{dispersion2}
    \left[\Omega-\Omega_0-{\bf U}_1\cdot({\bf k}-{\bf k}_0)\right]\left[\Omega-\Omega_0-{\bf U}_2\cdot({\bf k}-{\bf k}_0)
    \right]=\epsilon.
\end{equation}
Here ${\bf U}_{1,2}=\partial\Omega_{1,2}/\partial{\bf k}$ are the group velocities of the two modes calculated at a point
${\bf k}_0$ of the intersection line $\Omega_0({\bf k})$ (the latter is determined by the shape of the crossing modes) and
$\epsilon$ characterizes the coupling. From Eq.~(\ref{dispersion2}) one readily concludes that the effect of the coupling
critically depends on the sign of $\epsilon$. This is illustrated in Fig.~\ref{fig2}: When $\epsilon<0$ the crossing modes
merge and form a complex hybrid mode in a narrow range $|{\bf k}-{\bf k}_0|\propto\sqrt{\epsilon}$, with Re~$\Omega_{\rm
hyb}\simeq \Omega_0+\frac12({\bf U}_1+{\bf U}_2)\cdot({\bf k}-{\bf k}_0)$ and Im~$\Omega_{\rm hyb}\sim\pm\sqrt{\epsilon}$.
Thus, one of the branches of the hybrid mode has a positive imaginary part and hence can be unstable. When $\epsilon>0$ the
coupling is accompanied by reconnection (formation of a ``forbidden band'') and the modes remain stable (real). Furthermore,
the sign of the product ${\bf U}_1\cdot{\bf U}_2$ in the unstable case determines the type of the instability, which is
absolute for ${\bf U}_1\cdot{\bf U}_2<0$ (the case illustrated in Fig.~\ref{fig2}) and convective otherwise.

Now we can consider the DL modes described by Eqs (\ref{matrix}) and (\ref{dispersion}), with the dispersion and coupling
elements from Eqs (\ref{elements1}) and (\ref{elements2}). The detailed analysis shows that the intersection of the
transverse in-plane mode $\Omega_{{\rm h}\perp}({\bf k})$ with the longitudinal in-plane, $\Omega_{{\rm h}\|}({\bf k})$, as
well as with the out-of-plane mode $\Omega_{\rm v}({\bf k})$ is characterized by positive coupling and therefore is
accompanied by reconnection illustrated in Fig.~\ref{fig2}c, so that the modes remain stable. Moreover, the coupling for
these two pairs of modes exactly disappears (Fig.~\ref{fig2}a) if ${\bf k}$ is parallel to one of the principal lattice axes
(i.e., when $\theta=0^{\circ}$ or $30^{\circ}$ in Fig.~\ref{fig1}a).

In contrast, for the remaining pair of $\Omega_{\rm v}({\bf k})$ and $\Omega_{{\rm h}\|}({\bf k})$ the coupling is {\it
always negative}. As illustrated in Fig.~\ref{fig2}b, this results in the formation of the {\it hybrid} mode with the
imaginary part Im~$\Omega_{\rm hyb}\sim\pm\tilde q\tilde\delta$ (by the order of magnitude). If damping is low enough, this
triggers the {\it mode-coupling instability}.

Thus, only the intersection between the out-of-plane and longitudinal in-plane modes is critical for stability of 2D plasma
crystals \footnote{For 2D plasma crystals, the ``trivial'' origin of the non-Hamiltonial behavior associated with neutral
gas friction can often be neglected, since the corresponding damping rate is typically one-to-two orders of magnitude
smaller that the Einstein frequency \cite{RevModPhys.81.1353}.}.

\begin{figure*}
\includegraphics[width=0.9\textwidth,clip=]{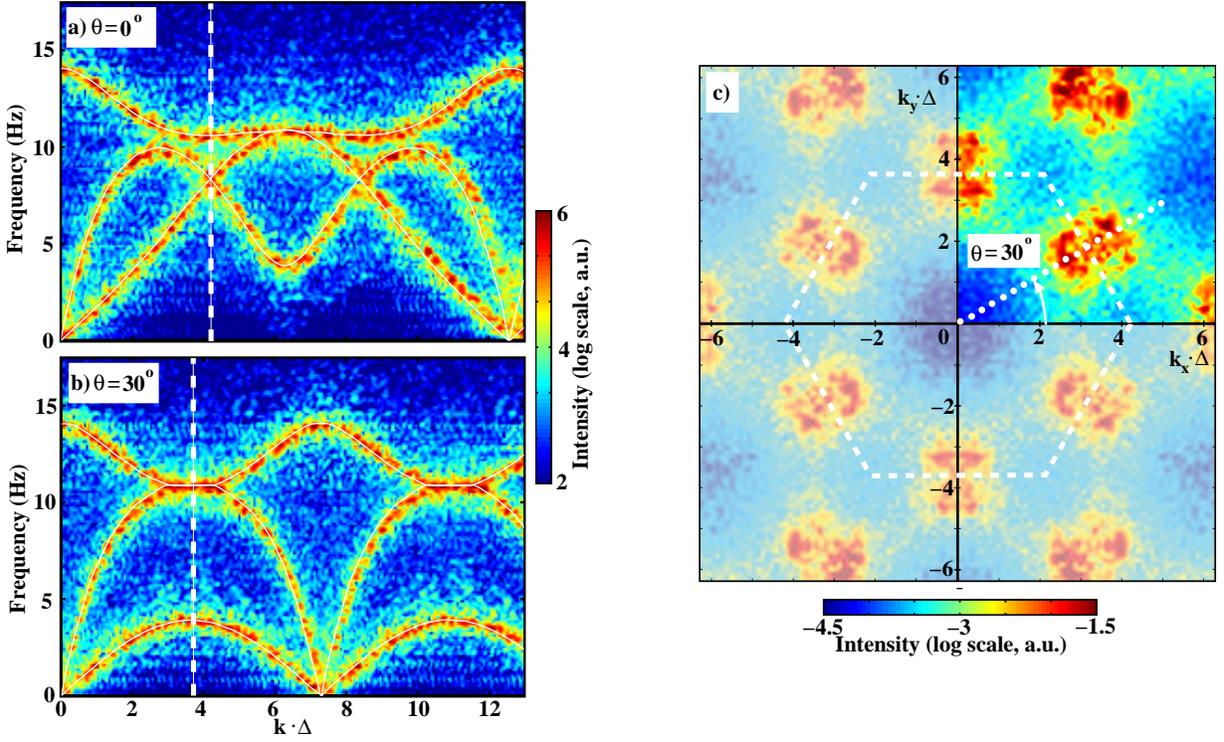}
\caption{Fluctuation spectra for ``shallow'' mode intersection (molecular dynamics simulations). Shown are the DL modes
(positive branches) for ${\bf k}$ at (a) $\theta=0^{\circ}$ and (b) $\theta=30^{\circ}$, and (c) spectrum in the ${\bf
k}$-plane integrated over frequency (in the range between 10~Hz~$\leq f\leq$~14~Hz). Simulations are for $N=16384$ particles
of a mass $m=6.1\times10^{-13}$~kg and charge $Q=-18500~e$, the screening length is $\lambda=600~\mu$m, the damping rate is
$\nu=0.87$~s$^{-1}$. The horizontal and vertical confinements have eigenfrequencies $f_{\rm conf}^{\rm(h)}=0.11$~Hz and
$f_{\rm conf}^{\rm(v)}=14.03$~Hz, respectively. The lattice constant in the center is $\Delta\simeq612~\mu$m
($\kappa=\Delta/\lambda\simeq1.02$). The wake parameters are $q=-0.2Q$ and $\delta=0.3\lambda$. The dashed lines show the
border of the first Brillouin zones, solid lines in (a) and (b) are theoretical curves. In (c) (as well as in other figures
depicting spectra in the ${\bf k}$ plane) the calculations for the upper right quadrant are mirrored in other three (shown
in lighter colors).}\label{fig3}
\end{figure*}

\subsection{Mode-coupling theory for 1D model}
\label{1D_theory}

It is instructive to present a brief summary of the mode-coupling theory which was originally developed for a 1D string
assuming the nearest-neighbor (NN) interactions \cite{PhysRevE.68.026405,PhysRevE.63.016409}. As it follows from the
subsequent analysis (see Sec.~\ref{Fingerprints}), this approach yields very simple expressions which can be conveniently
implemented for practical purposes (e.g., to calculate the instability thresholds, see Sec.~\ref{threshold}).

In the framework of 1D NN model \cite{PhysRevE.63.016409}, the coupling between horizontal (``longitudinal in-plane'') and
vertical (``out-of-plane'') DL modes,
\begin{equation}\label{Modes_1D}
    \begin{array}{l}
        \Omega_{{\rm h}}^2(k)=4(1-\tilde q)\left(\kappa^{-1}+2\kappa^{-2}+2\kappa^{-3}\right){\rm e}^{-\kappa}\sin^2\frac12k,\\[.3cm]
        \Omega_{{\rm v}}^2(k)=\Omega_{\rm conf}^2-4(1-\tilde q)\left(\kappa^{-2}+\kappa^{-3}\right){\rm e}^{-\kappa}\sin^2\frac12k,
    \end{array}
\end{equation}
is described by the following (exact) equation:
\begin{equation}\label{Coupling_1D}
    (\Omega^2-\Omega_{{\rm h}}^2)(\Omega^2-\Omega_{\rm v}^2)+\sigma^2=0.
\end{equation}
The range of $k$ is limited by the first Brillouin zone $|k|\leq\pi$ (we use the same normalization as above), and the 1D
coupling coefficient is
\begin{equation}\label{Coupling_coefficient_1D}
    \sigma(k)=2\tilde q\tilde\delta\left(\kappa^{-1}+3\kappa^{-2}+3\kappa^{-3}\right){\rm e}^{-\kappa}\sin k.
\end{equation}
All characteristics of the hybrid mode (e.g., width, growth rate, etc.) can be easily derived from Eqs
(\ref{Modes_1D})-(\ref{Coupling_coefficient_1D}), see Appendix~\ref{appendix1a}.

These results are obtained in the limit of ``small'' $\tilde\delta$ (for general case, see Ref. \cite{PhysRevE.63.016409}).
In Sec.~\ref{threshold} they will be compared with the exact 2D results derived in this limit.

\subsection{Additional remarks}
\label{remarks}

To conclude this Section, we should make two specific remarks which might be useful when considering DL waves and their
coupling:

(i) The dynamical matrix $\textsf{D}$ is solely characterized by interparticle interactions, i.e., does not depend on the
damping rate $\nu$. Thus, the {\it wave modes} $\Omega^2({\bf k})$ -- which are the eigenvalues of $\textsf{D}$ -- do not
depend on $\nu$ either. As regards the {\it dispersion relations} $\omega({\bf k})$, their dependence on $\nu$ is determined
by the solutions of the quadratic equation $\omega(\omega+ i\nu)-\Omega^2=0$. Therefore the effect of friction on waves is
straightforward: For weakly damped waves (when $|\Omega|\gg\nu$, which is typical for experiments), one readily obtains
$\omega({\bf k})\simeq\Omega({\bf k})-\frac12i\nu$. From the practical point of view this implies that one can analyze
undamped dispersion relations (i.e., put $\omega=\Omega$) and afterwards simply add $-\frac12\nu$ to the resulting imaginary
part \footnote{The real part of the dispersion relations remains unaffected by damping. This is only violated for the
acoustic modes $\Omega_{{\rm h}\|,\perp}({\bf k})$ in a close proximity of ${\bf k}={\bf 0}$ and therefore is unimportant
for the analysis.}, which simultaneously identifies the instability threshold for the hybrid mode \footnote{This also shows
that the analysis of wave modes does not require any distinction in terms of friction, as it has been done by Liu {\it et
al.}~\cite{PhysRevLett.105.085004} (where our previous theoretical results
\cite{PhysRevE.63.016409,PhysRevE.68.026405,zhdanov:zim} were erroneously referred to as ``frictionless'').} (see
Sec.~\ref{threshold2}).

(ii) For the analysis of wave modes in a crystal it is {\it sufficient} to consider the wave vectors from within the first
Brillouin zone \cite{kittel1961,lifshitz1981} which is shown in Fig.~\ref{fig1}b. This zone is nothing but the
Wigner-Seitz cell of the reciprocal lattice formed by the basis vectors ${\bf b}_{1}$ and ${\bf b}_{2}$. Hence, the wave
vectors ${\bf k}$ and ${\bf k}'={\bf k}+{\bf G}$ which are different by a linear combination of the basis vectors (${\bf
G}=m{\bf b}_1+n{\bf b}_2$) are equivalent for wave modes, i.e., $\Omega({\bf k}+{\bf G})\equiv\Omega({\bf k})$.

\begin{figure}
\includegraphics[width=0.8\columnwidth,clip=]{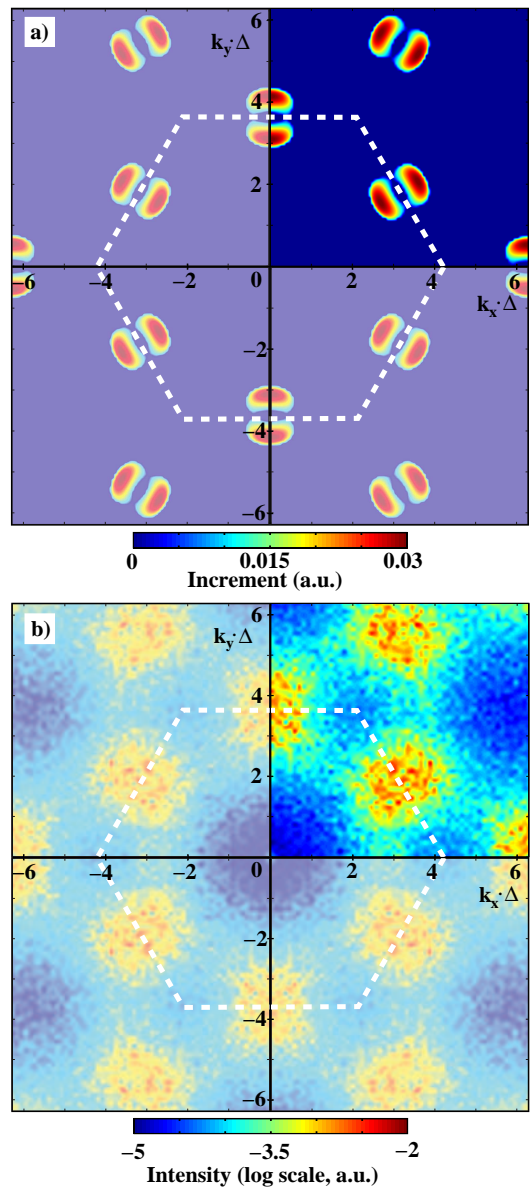}
\caption{(a) Contour plot of the growth rate Im~$\omega_{\rm hyb}({\bf k})$ calculated from theory and (b) fluctuation
spectra in the ${\bf k}$-plane obtained from simulations (integrated over frequency in the range between 10~Hz~$\leq
f\leq$~14~Hz). To facilitate comparison with theory, the simulations are performed with periodic boundary conditions
in the horizontal plane, otherwise the parameters are the same as in Fig.~\ref{fig3}. }\label{fig4}
\end{figure}

One has to remember this fact when analyzing dispersion relations at large $|{\bf k}|$. For instance, recently Liu {\it et
al.}~\cite{PhysRevLett.105.085004} identified one of the ``hot spots'' seen in their fluctuation spectra (in their Fig.~4c)
as a new hybrid mode. This spot is centered at $|{\bf k}|\Delta\simeq6.3(\simeq2\pi)$, $\theta=0^{\circ}$ and therefore is
located outside the first Brillouin zone; after the mapping with, say, ${\bf G}={\bf b}_2$, its position is $|{\bf
k}|\Delta\simeq \frac2{\sqrt{3}}\pi$, $\theta=90^{\circ}$ (equivalent to $30^{\circ}$, see Fig.~\ref{fig1}b; note also the
change in the polarization, from the transverse to longitudinal). This naturally coincides with the position of the
``regular'' hybrid mode at the border of the first zone, whereas the ``original'' spot is merely its image outside it (as
illustrated in Figs~\ref{fig3}c and \ref{fig4}b).

\section{Fingerprints of Mode Coupling}
\label{Fingerprints}

In this Section we present detailed discussion and analysis of distinct fingerprints characterizing the wake-mediated
coupling of DL modes in 2D crystals. Based on the results of Sec.~\ref{types} we naturally focus only on the coupling
between the out-of-plane and longitudinal in-plane modes, which causes the formation of the hybrid mode and can trigger the
mode-coupling instability.

To illustrate key features of the mode coupling, the theoretical calculations were combined with the results of molecular
dynamics (MD) simulations (see description in Appendix~\ref{appendix2}). The DL wave modes were obtained from the
simulations by plotting the so-called ``fluctuation spectra'' -- intensity distributions of thermal velocity fluctuations in
the $(\omega,{\bf k})$ space.

\subsection{Hybrid mode}\label{hybrid}

Figure~\ref{fig3} represents a characteristic example of fluctuation spectra measured at the onset of the mode-coupling
instability. It was obtained from MD simulations performed for the conditions close to those of experiment by Cou\"edel {\it
et al.}~\cite{PhysRevLett.104.195001} (experiment II in Table~\ref{tab1}). In this case the out-of-plane and longitudinal
in-plane modes are just barely touching. Such ``shallow'' intersection is nevertheless sufficient to form a distinct hybrid
mode and to trigger the instability.

Figures~\ref{fig3}a and b depict the DL modes in the $(\omega,|{\bf k}|)$ plane (positive branches). They are obtained for
two directions of ${\bf k}$ at $\theta=0^{\circ}$ and $\theta=30^{\circ}$, respectively (corresponding to the principal
lattice axes, see Fig.~\ref{fig1}a). Figure~\ref{fig3}c shows the intensity distribution in the ${\bf k}$-plane (integrated
over frequency), which provides a ``top view'' on the fluctuation spectra (dotted line shows the direction of
$\theta=30^{\circ}$). The hybrid mode can be identified in Figs~\ref{fig3}b and c as ``hot spots'' located near the border
of the first Brillouin zone, where the intersection occurs (note also images of the hybrid mode outside the first zone). The
theoretical curves calculated from Eqs (\ref{matrix}), (\ref{dispersion}), (\ref{elements1}), and (\ref{elements2}) for the
same set of parameters demonstrate excellent agreement with simulations.

The reason why the fluctuation intensity is anomalously high at the position of the hybrid mode is the heating induced by
the wake-mediated coupling. Figure~\ref{fig3} represents the marginally unstable regime, when Im~$\omega_{\rm
hyb}(\simeq{\rm Im}~\Omega_{\rm hyb}-\frac12\nu)\gtrsim0$. In this regime (also observed in experiment
\cite{PhysRevLett.104.195001}) the effective growth rate of the instability is low, which allows us to obtain the
fluctuation spectra before the crystal is eventually destroyed. It is worth noting, however, that the hybrid mode would
usually appear as a ``hot spot'' even if the instability is suppressed by friction and Im~$\omega_{\rm hyb}<0$. This is
because the heating due to mode coupling is present also in stable regime. It becomes negligible only when Im~$\Omega_{\rm
hyb}\ll\frac12\nu$, and then the fluctuation intensity coincides with the intensity of DL modes outside the hybrid zone
(where it is determined by the background neutrals or temperature of thermostat in simulations).

At the initial stage of the instability (and, of course, when the instability is suppressed) the effect of nonlinearity on
the DL modes is not significant. In this case the contour plot of the growth rate Im~$\omega_{\rm hyb}({\bf k})$ reasonably
reproduces the distribution of fluctuation intensity in the vicinity of the hybrid mode. This is illustrated in
Fig.~\ref{fig4} where we plotted Im~$\omega_{\rm hyb}({\bf k})$ predicted by theory and the (frequency-integrated)
fluctuation spectra from simulations. At a later stage, however, a nonlinear coupling between different modes becomes
essential. This results in a variety of new phenomena, such as the energy cascades from the ``hot spots'' in the ${\bf k}$
space and the generation of secondary harmonics at the double frequency of the hybrid mode (which are clearly seen in Fig.~3
of Ref.~\cite{PhysRevLett.104.195001}). Nevertheless, since the principal aim of this paper is to focus on distinct features
of the {\it linear mode coupling}, we leave aside the discussion of numerous nonlinear coupling effects.

For the plots shown in Fig.~\ref{fig4} we used the same set of parameters as for Fig.~\ref{fig3}, but for simulations we
assumed periodic boundary conditions. The latter not only allowed us to directly compare the theory and simulations, but
also to probe the effect of weak deviation from periodicity caused by the horizontal confinement. We see that the only
noticeable difference between Figs~\ref{fig3}c and \ref{fig4}b is the intensity of the hybrid mode. This reflects the fact
that the growth rate of the instability critically depends on the interparticle distance $\Delta$ (resulting in
exponentially growing difference in the heating at the initial stage). Therefore in the case of parabolic confinement (where
$\Delta$ in the center is somewhat smaller than the average value) the intensity of the hybrid mode is higher than in the
homogeneous case shown in Figs~\ref{fig4}b. Otherwise, the positions and the overall shape of the ``hot spots'' in these two
plots practically coincide, suggesting that weak inhomogeneity which is always present in experiments due to the horizontal
confinement does not really affect the periodic structure of the Brillouin zones.

\begin{figure}
\includegraphics[width=0.8\columnwidth,clip=]{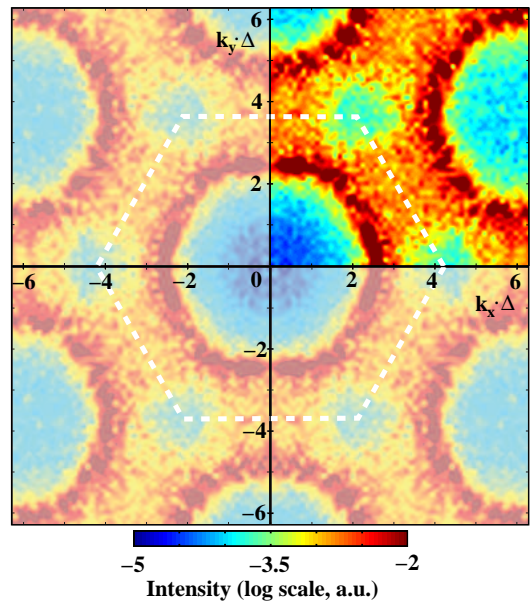}
\caption{Illustration of ``deep'' mode intersection (simulations). The eigenfrequency of the vertical confinement is
$f_{\rm conf}^{\rm(v)}=13.12$~Hz (spectra are integrated over frequency in the range between 9~Hz~$\leq f\leq$~13~Hz), the
damping rate is $\nu=2.19$~s$^{-1}$. Otherwise the parameters are the same as in Fig.~\ref{fig3}.}\label{fig5}
\end{figure}

Finally, to illustrate the difference between ``shallow'' and ``deep'' intersections of the out-of-plane and longitudinal
in-plane modes, in Fig.~\ref{fig5} we demonstrate the fluctuation spectra obtained for the eigenfrequency of vertical
confinement $f_{\rm conf}^{\rm(v)}$ which is slightly lower than that in Fig.~\ref{fig3}. This naturally shifts the
out-of-plane mode down and thus makes the resulting hybrid mode much more pronounced: Figure~\ref{fig5} shows that for deep
intersections the separate ``islands'' of the hybrid mode merge into a joint broad ``ring'' (within the first Brillouin
zone) and the ``hot'' region occupies substantially larger part of the ${\bf k}$-plane. The magnitude of Im~$\Omega_{\rm
hyb}$ for deep intersections is usually much larger than for the shallow intersections and therefore the threshold
conditions to trigger the mode-coupling instability in the former case are substantially relaxed (see discussion in
Sec.~\ref{threshold}). Note that in this case, in order to keep the crystal in the marginally unstable regime (which allows
us to recover the fluctuation spectra) we increased the damping rate to $\nu=2.19$~s$^{-1}$.

\subsection{Angular dependence}\label{angular}

The dependence of DL modes on the direction of ${\bf k}$ is evident from the discussion above. For the hybridization and the
onset of the mode-coupling instability this dependence becomes crucial and therefore it deserves a separate discussion.

Figures~\ref{fig3} and \ref{fig4} show that the first crossing of the out-of-plane and longitudinal in-plane modes always
occurs at $\theta\simeq30^{\circ}$, at the border of the first Brillouin zone. The area of the hybrid mode in the ${\bf k}$
plane rapidly grows as the intersection gets deeper, and eventually the hybrid mode exists for any $\theta$, as in
Fig.~\ref{fig5}. The crossing depth is controlled by changing the particle density or/and the eigenfrequency of the vertical
confinement (see Sec.~\ref{threshold}).

Therefore, when the DL fluctuation spectra are measured in experiments, under conditions close to the onset of hybridization
(and, if not suppressed -- of the mode-coupling instability), special attention should be paid to the orientation of ${\bf
k}$ with respect to the crystalline lattice. Under these circumstances the direction should be chosen according to the
purpose of the experiment: If detection of the mode coupling is the goal, then $\theta=30^{\circ}$ is the best choice; if
unperturbed modes are to be measured, then $\theta=0^{\circ}$ should be used. The experimental technique must therefore
provide angular resolution, which is easily done for the in-plane modes, but is not always possible for the out-of-plane
mode (since it is difficult to achieve angular positioning of the plasma crystal in an experiment). For instance, if the
out-of-plane mode is measured with a fixed side-view camera~\cite{PhysRevLett.105.085004} then the angle $\theta$ is set
during the experiment and cannot be adjusted at the analysis stage, but if it is measured with a top-view camera (from the
variation of individual particle intensity)~\cite{couedel:215001} then $\theta$-resolved wave spectra can be obtained.

\subsection{Mixed polarization}

\begin{figure}[h]
\includegraphics[width=0.92\columnwidth]{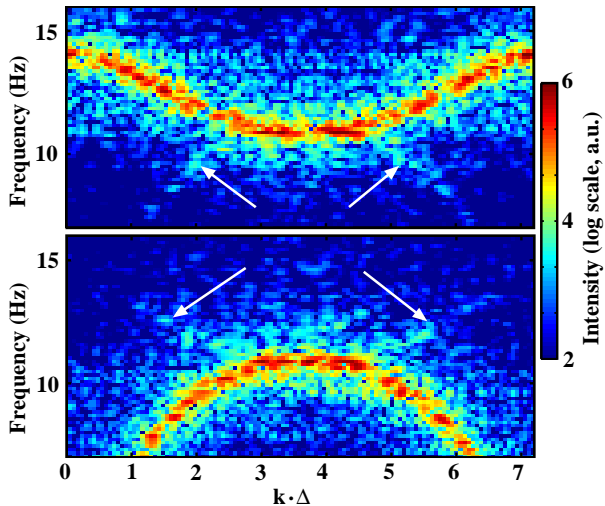}
\caption{Mixed polarization near the hybrid mode (simulations). Shown are the fluctuation spectra with the transverse
out-of-plane (a) and longitudinal (b) polarization, for ${\bf k}$ at $\theta=30^{\circ}$ and parameters of Fig.~\ref{fig3}.
Arrows indicate traces of mixed polarization. Far from the hybrid mode (``hot spots'') the modes become purely out-of-plane
(a) and in-plane (b).}\label{fig6}
\end{figure}

As we pointed out in Sec.~\ref{theory}, the DL wave modes are independent only when the wake-mediated coupling can be
neglected: Far from the intersection or if the modes do not cross at all, the out-of-plane mode is purely transverse, one of
the in-plane modes is longitudinal and the other is transverse. However, near the hybrid mode, where the coupling between
the first two modes is strong, the theory predicts the emergence of {\it mixed polarization}\cite{zhdanov:zim}: As the modes
approach the junction point, their polarizations become {\it oblique} (in the plane formed by ${\bf k}$ and the vertical
axis), but the eigenvectors remain {\it mutually orthogonal}. At the point where they merge the polarization exhibits a
discontinuity -- the eigenvectors of the two modes form a single (hybrid) eigenvector with the elliptic polarization.

Traces of the mixed polarization for the out-of-plane mode have been observed in our recent
experiment~\cite{PhysRevLett.104.195001} (supplemental Fig.~S2 therein) as well as in the experiment by Liu {\it et
al.}~\cite{PhysRevLett.105.085004} (shown in their Fig.~4b and erroneously interpreted as a ``new mode''). In
Fig.~\ref{fig6} we illustrate the mixed polarization observed in simulations (performed for the conditions of our
experiment~\cite{PhysRevLett.104.195001}). The magnitude of this effect ($\propto\tilde q\tilde\delta<1$) is quite small for
typical experimental conditions, so that its detection requires very careful analysis and high signal-to-noise ratio.

\subsection{Thresholds}\label{threshold}

The mode-coupling theory predicts two distinct thresholds: (i) {\it Confinement threshold}, which determines when the
out-of-plane and longitudinal in-plane modes intersect; it depends on the combination of the vertical confinement frequency
and the lattice constant (particle number density). This threshold identifies the sufficient condition for the hybrid mode
to form, but only the necessary condition for the instability to set in. The sufficient condition for the instability is
determined by (ii) {\it damping threshold} -- this identifies the critical value of the damping rate (controlled by
pressure) below which the instability is not suppressed. Let us discuss thresholds (i) and (ii) separately (we recall that
all frequencies are normalized by $\Omega_{\rm DL}$).

\begin{figure}
\includegraphics[width=7.5cm,clip=]{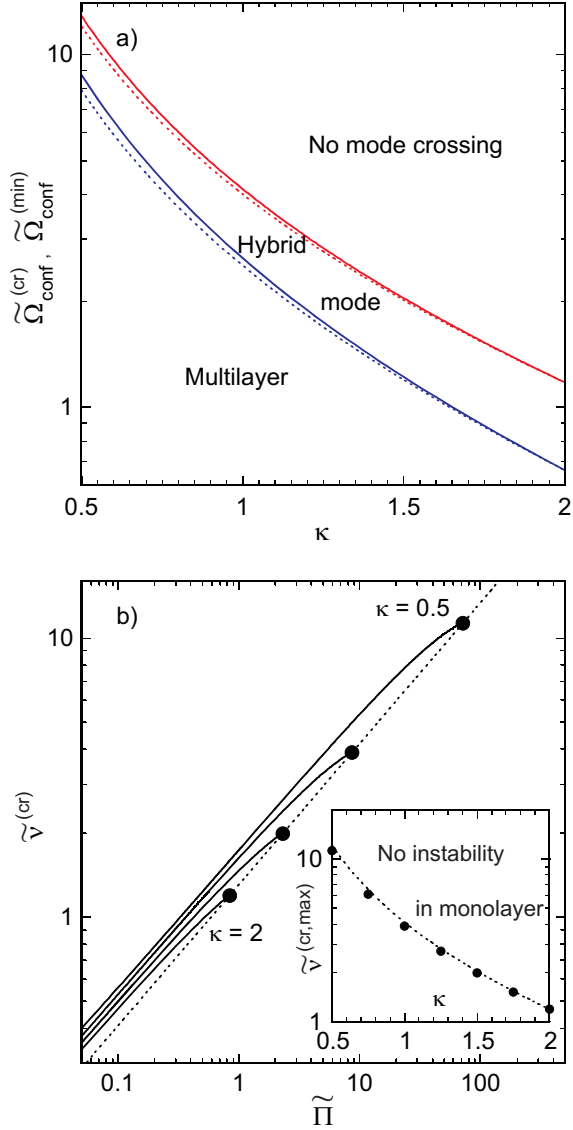}
\caption{Universal mode-coupling thresholds (limit of small $\tilde\delta$). (a) Critical value of the vertical confinement frequency,
$\Omega_{\rm conf}^{\rm(cr)}$ (confinement threshold, upper curves), which identifies the formation of the hybrid mode, shown
versus the screening parameter $\kappa=\Delta/\lambda$. Also plotted is the minimum value $\Omega_{\rm conf}^{\rm(min)}$
(below which only multilayer crystals can exist, lower curves). (b) Critical value of the damping rate, $\nu^{\rm(cr)}$ (damping
threshold), which determines the sufficient condition for the instability to set in, shown versus the confinement control
parameter, $\Pi=(\Omega_{\rm conf}^{\rm(cr)})^2 -\Omega_{\rm conf}^2$, for $\kappa=0.5$, 1, 1.5, and 2. The upper bound for the
critical damping rate, $\nu^{\rm(cr,max)}$, is reached when $\Omega_{\rm conf}=\Omega_{\rm conf}^{\rm(min)}$
(marked by bullets). The inset in (b) shows $\nu^{\rm(cr,max)}$ for different $\kappa$, the mode-coupling instability is always
suppressed above the dotted line. In both panels, solid lines are exact 2D results derived in this paper, dashed lines represent
results of a 1D NN theory\cite{PhysRevE.63.016409}. Tilde denotes universal normalization (see text for details).}\label{fig7}
\end{figure}

\subsubsection{Confinement threshold}\label{threshold1}

The first intersection of the out-of-plane and longitudinal in-plane modes takes place at the border of the Brillouin zone
\cite{zhdanov:zim}, for ${\bf k}$ at $\theta=30^{\circ}$. Then, using the expressions for $\Omega_{\rm v}({\bf k})$ and
$\Omega_{{\rm h}\|}({\bf k})$ we readily obtain the condition for the hybrid mode to form (confinement threshold),
\begin{equation}\label{threshold_confinement1}
    \Omega_{\rm conf}<\Omega_{\rm conf}^{\rm(cr)}=\left.\sqrt{\alpha_{\rm h}+2\alpha_{\rm v}+|\beta|}\right|_{{\bf k}
    ={\bf k}_{\rm cr}},
\end{equation}
where $|{\bf k}_{\rm cr}|=\frac2{\sqrt{3}}\pi$.

\begin{figure}
\includegraphics[width=8cm,clip=]{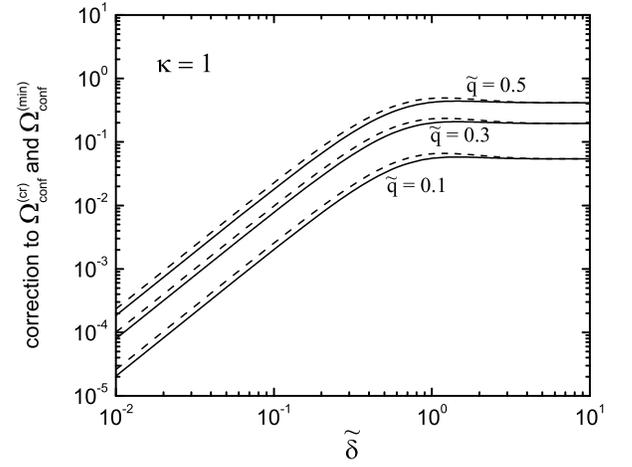}
\caption{Role of ``finite'' $\tilde\delta$. Shown is the relative deviation of $\Omega_{\rm conf}^{\rm (cr)}$ (solid line) and
$\Omega_{\rm conf}^{\rm (min)}$ (dashed line) from the values calculated in the limit of small $\tilde\delta$
(see Fig.~\ref{fig7}a). The results are illustrated for $\kappa=1$ and three different values of $\tilde q$.}\label{fig8}
\end{figure}

Generally, the critical confinement frequency $\Omega_{\rm conf}^{\rm(cr)}$ is a rather complicated function of $\kappa$,
$\tilde\delta$, and $\tilde q$; it is given in Appendix~\ref{appendix3}, Eq. (\ref{app_confinement1}). However, Eq.
(\ref{elements1}) suggests that to the accuracy $O(\tilde\delta)$ the r.h.s. of Eq. (\ref{threshold_confinement1}) is
independent of $\tilde\delta$ and is proportional to $\sqrt{1-\tilde q}$. Thus, in this limit of ``small'' $\tilde\delta$
the combination $\Omega_{\rm conf}^{\rm(cr)}/\sqrt{1-\tilde q}\equiv\widetilde\Omega_{\rm conf}^{\rm(cr)}$ is a function of
$\kappa$ only (i.e., of the particle density) which is given by Eq. (\ref{app_confinement2}). Figure~\ref{fig7}a shows this
universal dependence in the useful range $0.5<\kappa<2$ (upper solid line).

One should always keep in mind that $\Omega_{\rm conf}$ cannot be arbitrarily small
\cite{PhysRevE.63.016409,PhysRevE.68.026405}: The out-of-plane mode $\Omega_{\rm v}^2({\bf k})= \Omega_{\rm
conf}^2-2\alpha_{\rm v}$ decreases monotonically with $|{\bf k}|$ and attains a minimum at the border of the first Brillouin
zone. If $\Omega_{\rm v}^2<0$ this formally implies instability. Physically, this is because the vertical confinement in
this case becomes too weak to keep the particles in the monolayer and the bifurcation to a multilayer crystal occurs (i.e.,
a monolayer crystal is no longer a ground state). Thus, irrespective of the value of damping rate, a monolayer plasma
crystal can only exist if the vertical confinement frequency exceeds the minimum value defined by \footnote{$\Omega_{\rm
v}^2({\bf k})$ is practically independent of $\theta$ at the border of the Brillouin zone \cite{zhdanov:zim} (the variation
is $\lesssim1~\%$), so that in order to facilitate the calculations we substitute ${\bf k}$ at $\theta=30^\circ$.}
\begin{equation}\label{threshold_confinement3}
    \Omega_{\rm conf}>\Omega_{\rm conf}^{\rm(min)}\simeq\left.\sqrt{2\alpha_{\rm v}}\right|_{{\bf k}={\bf k}_{\rm cr}}.
\end{equation}
This value is given by Eqs (\ref{app_confinement3}) and (\ref{app_confinement4}) for arbitrary and small values of
$\tilde\delta$, respectively. In the latter case we obtain a universal dependence $\Omega_{\rm conf}^{\rm(min)}/
\sqrt{1-\tilde q}\equiv\widetilde \Omega_{\rm conf}^{\rm(min)}$ on $\kappa$ which is also plotted in Fig.~\ref{fig7}a (lower
solid line).

The effect of ``finite'' $\tilde\delta$ [i.e., beyond the accuracy $O(\tilde\delta)$] is illustrated in Fig.~\ref{fig8} for
$\kappa=1$ (results are practically independent of $\kappa$ in the range considered here). We plot the {\it relative
deviations} of $\Omega_{\rm conf}^{\rm (cr)}$ and $\Omega_{\rm conf}^{\rm (min)}$ from the values given by Eqs
(\ref{app_confinement2}) and (\ref{app_confinement4}), respectively. The deviations increase as $\propto\tilde\delta^2$ for
$\tilde\delta\lesssim0.3$ and then rapidly saturate at the level $(1-\tilde q)^{-1/2}-1$. [The latter is because both
$\Omega_{\rm conf}^{\rm (cr)}$ and $\Omega_{\rm conf}^{\rm (min)}$ do not depend on the wake parameters for
$\tilde\delta\gg1$, whereas at $\tilde\delta\ll1$ they scale as $\propto\sqrt{1-\tilde q}$.] Note that the curves have
almost identical shape at different $\tilde q$, which indicates that the shown correction approximately scales as
$\propto(1-\sqrt{1-\tilde q})$. Given typical experimental errors (see Table~\ref{tab1} and Fig.~\ref{fig10}), we conclude
that the corrections shown in Fig.~\ref{fig8} can generally be neglected, i.e., the accuracy of universal dependencies
obtained in the limit of small $\tilde\delta$ is sufficient for practical purposes. In some cases, however, the effect of
``finite'' $\tilde\delta$ can nevertheless play an important role (see Sec.~\ref{large_delta}).

Finally, let us compare the ``exact'' 2D results (in the limit of small $\tilde\delta$) with the results of approximate 1D
NN mode-coupling theory summarized in Sec.~\ref{1D_theory}. For the 1D case, the critical and minimum confinement
frequencies are derived from Eq.~(\ref{Modes_1D}) using the conditions $\Omega_{\rm h}^2(\pi)=\Omega_{\rm v}^2(\pi)$ and
$\Omega_{\rm v}^2(\pi)=0$, respectively. For the comparison with the 2D results, we multiply the 1D results by constant
``form factors'' reflecting mismatch of 1D and 2D environments, which yields
\begin{equation}\label{threshold_confinement_1D}
    \begin{array}{l}
        \widetilde\Omega_{\rm conf}^{\rm(cr)} \simeq 2F_{\rm conf}^{\rm(cr)}\sqrt{\left(\kappa^{-1}+3\kappa^{-2}
        +3\kappa^{-3}\right){\rm e}^{-\kappa}},\\[.3cm]
        \widetilde\Omega_{\rm conf}^{\rm(min)} \simeq 2F_{\rm conf}^{\rm(min)}\sqrt{\left(\kappa^{-2}+\kappa^{-3}\right)
        {\rm e}^{-\kappa}}.
    \end{array}
\end{equation}
The form factors $F_{\rm conf}^{\rm(cr)}=1.25$ and $F_{\rm conf}^{\rm(min)}=1.49$ are determined from a fit to the 2D
results, Eqs (\ref{app_confinement2}) and (\ref{app_confinement4}). The dependencies given by
Eq.~(\ref{threshold_confinement_1D}) are shown in Fig.~\ref{fig7}a by dashed lines.

Thus, we see that the 1D NN theory provides excellent qualitative description of the confinement thresholds -- even for
small $\kappa$; by multiplying these results with {\it constant} factors about unity we also get very good quantitative
agreement with 2D results.

\subsubsection{Damping threshold}\label{threshold2}

As pointed out in Sec.~\ref{types}, for $\theta=30^{\circ}$ (as well as for $0^{\circ}$) the transverse in-plane mode
becomes exactly decoupled. In this case the out-of-plane and longitudinal in-plane modes are governed by the following
equation [similar to that for the 1D NN model, Eq.~(\ref{Coupling_1D})]:
\begin{equation*}
    (\Omega^2-\Omega_{{\rm h}\|}^2)(\Omega^2-\Omega_{\rm v}^2)+\sigma_{y}^2=0.
\end{equation*}
In the vicinity of the intersection $(\Omega_0,{\bf k}_0)$ the coupled modes are described by Eq. (\ref{dispersion2}) with
$\epsilon\simeq(\sigma_y/2\Omega_0)^2$. The group velocities ${\bf U}_{{\rm h}\|}$ and ${\bf U}_{\rm v}$ are collinear with
${\bf k}$ due to the azimuthal symmetry at $\theta=30^{\circ}$, so that the hybrid mode is given by
\begin{eqnarray*}
    \Omega_{\rm hyb}({\bf k})=\Omega_0+\frac12({\bf U}_{{\rm h}\|}+{\bf U}_{\rm v})\cdot({\bf k}-{\bf k}_0)\hspace{2cm}\\
    \pm\frac12i\sqrt{\sigma_y^2({\bf k}_0)/\Omega_0^2-|{\bf U}_{{\rm h}\|}-{\bf U}_{\rm v}|^2|{\bf k}-{\bf k}_0|^2}.
\end{eqnarray*}
Thus, the width of the hybrid mode is $|{\bf k}-{\bf k}_0|=\sigma_y({\bf k}_0)/(\Omega_0|{\bf U}_{{\rm h}\|}-{\bf U}_{\rm
v}|)^{-1}$, and the maximum of Im~$\Omega_{\rm hyb}({\bf k})$ is attained at the crossing point ${\bf k}_0$. From this we
immediately obtain the sufficient condition for the instability (damping threshold),
\begin{equation}\label{threshold_pressure1}
    \nu<\nu^{\rm(cr)}=\sigma_y({\bf k}_0)/\Omega_0,
\end{equation}
where $\Omega_0$ and ${\bf k}_0$ are determined from the crossing condition,
\begin{equation}\label{threshold_pressure1a}
    \Omega_0=\left.\sqrt{\alpha_{\rm h}+\beta}\right|_{{\bf k}={\bf k}_{0}}=\left.\sqrt{\Omega_{\rm conf}^2-2\alpha_{\rm
    v}}\right|_{{\bf k}={\bf k}_{0}}.
\end{equation}
In order to relate Eq. (\ref{threshold_pressure1}) to the necessary instability condition [confinement threshold, Eq.
(\ref{threshold_confinement1})] we introduce the {\it confinement control parameter},
\begin{displaymath}
    \Pi=(\Omega_{\rm conf}^{\rm(cr)})^2-\Omega_{\rm conf}^2,
\end{displaymath}
which varies in the range from zero to $(\Omega_{\rm conf}^{\rm(cr)})^2-(\Omega_{\rm conf}^{\rm(min)})^2$. It is obtained
from Eq. (\ref{threshold_pressure1a}),
\begin{equation}\label{threshold_pressure2}
    \Pi=\left.\left(\alpha_{\rm h}+2\alpha_{\rm v}+\beta\right)\right|_{{\bf k}_0}^{{\bf k}_{\rm cr}},
\end{equation}
where the r.h.s. is calculated as the difference at ${{\bf k}_{\rm cr}}$ and ${{\bf k}_0}$. Thus, Eqs
(\ref{threshold_pressure1}) and (\ref{threshold_pressure2}) determine the dependence of the critical damping rate
$\nu^{\rm(cr)}$ on the control parameter $\Pi$ via the parametric dependence on $|{\bf k}_0|$. For arbitrary $\tilde\delta$,
this dependence is derived in Appendix~\ref{appendix3}, Eqs (\ref{app_pressure1}) and (\ref{app_pressure2}).

Similar to the confinement threshold discussed in the previous subsection, for the damping threshold one can also obtain a
universal dependence in the limit of small $\tilde\delta$. It is calculated in Eqs~(\ref{app_pressure3}) and
(\ref{app_pressure4}), where the r.h.s. are functions of $\kappa$ only (along with the parameter $|{\bf k}_0|$) and the
dependence on the wake parameters is explicitly given by the l.h.s. By introducing the universal normalization for the
critical damping rate and the confinement control parameter,
\begin{equation}\label{universal_norm}
    \tilde\nu^{\rm(cr)}\equiv\frac{\sqrt{1-\tilde q}}{\tilde q\tilde\delta}\nu^{\rm(cr)},\quad
    \widetilde\Pi\equiv\frac{\Pi}{1-\tilde q},
\end{equation}
we conclude that $\tilde\nu^{\rm(cr)}$ is a universal function of $\kappa$ and $\widetilde\Pi$. Figure~\ref{fig7}b shows
$\tilde\nu^{\rm(cr)}$ versus $\widetilde\Pi$ for four characteristic values of $\kappa$ (solid lines). For ``shallow''
intersections, when $\Omega_{\rm conf}$ is slightly below the critical value $\Omega_{\rm conf}^{\rm(cr)}(\kappa)$ the
critical damping rate increases as $\tilde\nu^{\rm(cr)}\propto \sqrt{\widetilde\Pi}$, which can be readily derived from
Eqs~(\ref{app_pressure3}) and (\ref{app_pressure4}). Thus, for given $\kappa$ and $\widetilde\Pi$ (with $\Omega_{\rm
conf}^{\rm(min)}< \Omega_{\rm conf}<\Omega_{\rm conf}^{\rm(cr)}$) monolayer plasma crystals are unstable if $\nu<(\tilde
q\tilde\delta/\sqrt{1-\tilde q})\tilde\nu^{\rm(cr)}(\kappa,\widetilde\Pi)$ and stable otherwise.

The critical damping rate is always bound from above, $\tilde\nu^{\rm(cr)}\leq \tilde\nu^{\rm(cr,max)}$. For a given
$\kappa$, the upper bound $\tilde\nu^{\rm(cr,max)}(\kappa)$ corresponds to $\Omega_{\rm conf}=\Omega_{\rm
conf}^{\rm(min)}(\kappa)$ and is marked by bullets in Fig.~\ref{fig7}b. These points shown in the inset as a function of
$\kappa$ identify the {\it absolutely stable} region: The mode-coupling instability in monolayer crystals is always
suppressed when $\nu$ exceeds $(\tilde q\tilde\delta/\sqrt{1-\tilde q})\tilde\nu^{\rm(cr,max)}(\kappa)$.

\begin{table*}[htbp]
\caption{Parameters of 2D crystals for different experiments (marked I to XII). The mode crossing is indicated as observed
in that study.}\label{tab1}
\begin{tabular}{lcccccccc|c|cc}
    \hline
    \hline
                                           & $m$             & $|Q|$ & $\Delta$  & $\lambda$  & $\kappa$  & $f_{\rm conf}^{\rm(v)}$ & $\nu$      & $\omega_0$   & Mode      & ~$2\pi f_{\rm conf}^{\rm(v)}/\omega_0$~ & ~$\nu/\omega_0$~ \\
                                           & ($10^{-13}$~kg) & ($e$) & ($\mu$m)  & ($\mu$m)   &           & (Hz)                    & (s$^{-1}$) & (s$^{-1}$)   & crossing  &                                         &                  \\
    \hline
I \cite{PhysRevLett.104.195001}$^{\ddag}$  & 6.1             & 17400 &  600      &  570       & 1.05      & 16.5                    & 0.9        & 23.0         &  No       & 4.5                                     & 0.04             \\
II \cite{PhysRevLett.104.195001}$^{\ddag}$ & 6.1             & 18300 &  640      &  610       & 1.05      & 14.5                    & 0.9        & 22.0         &  Yes      & 4.1                                     & 0.04             \\
III \cite{couedel:215001}$^{\ddag}$        & 5.3             & 14000 &  525      &  400       & 1.30      & 23.3                    & 1.0        & 24.3         &  No       & 6.0                                     & 0.04             \\
IV \cite{couedel:215001}$^{\ddag}$         & 5.3             & 14200 &  514      &  514       & 1.00      & 20.5                    & 1.0        & 25.4         &  No       & 5.1                                     & 0.04             \\
V \cite{couedel:215001}$^{\ddag}$          & 5.3             & 14600 &  1060     &  815       & 1.30      & 18.3                    & 1.0        & 8.8          &  No       & 13.0                                    & 0.11             \\
VI \cite{couedel:215001}$^{\ddag}$         & 5.3             & 18000 &  600      &  600       & 1.00      & 16.3                    & 1.1        & 25.6         &  Yes      & 4.0                                     & 0.04             \\
VII \cite{PhysRevLett.105.085004}          & 4.2             & 9000  &  1000     &  700       & 1.40      & 13.9                    & 3.5        & 6.7          &  No       & 13.0                                    & 0.52             \\
VIII \cite{PhysRevLett.105.085004}         & 4.2             & 9000  &  600      &  700       & 0.86      & 13.9                    & 3.5        & 14.4         &  Yes      & ~6.1$^{\dag}$                           & 0.24             \\
IX \cite{PhysRevLett.105.269901}           & 4.2             & 12000 &  600      &  760       & 0.79      & 13.9                    & 3.5        & 19.2         &  Yes      & 4.5                                     & 0.18             \\
X \cite{PhysRevE.68.026405}                & 5.5             & 15500 &  550      &  500       & 1.10      & 15.5                    & 3.3        & 24.6         &  Yes      & 4.0                                     & 0.14             \\
XI \cite{samsonov:026410}                  & 5.6             & 16000 &  700      &  930       & 0.75      & 16.0                    & 1.2        & 17.6         &  No       & 5.7                                     & 0.07             \\
XII \cite{samsonov:026410}                 & 5.6             & 13000 &  540      &  635       & 0.85      & 21.0                    & 2.2        & 21.1         &  No       & 6.3                                     & 0.10             \\
   \hline
   \hline
\end{tabular}\\
{$^{\dag}$ The value is too high for the mode crossing (see Fig.~\ref{fig10}), amended in Erratum \cite{PhysRevLett.105.269901}.}\\
{$^{\ddag}$ For these data, error bars are available: $\pm15~\%$ for $\kappa$ and $\lambda$, $\pm10~\%$ for $|Q|$, $\pm8~\%$ for $f_{\rm conf}^{\rm(v)}$, and $\pm4~\%$ for $\Delta$.}
\end{table*}

According to the 1D NN theory [see Appendix~\ref{appendix1a}, Eq. (\ref{hybrid_1D})] the critical damping rate is
\begin{equation*}
    \tilde\nu^{\rm(cr)}=\sqrt{\frac{\kappa^2+3\kappa+3}{\kappa^2+2\kappa+2}}\sqrt{\widetilde\Pi},
\end{equation*}
i.e., it also scales as a square root of the control parameter. The upper bound for $\nu^{\rm(cr)}(\kappa)$ is obtained by
substituting Eq. (\ref{threshold_confinement_1D}) for the maximum of $\Pi$, which yields
\begin{equation}\label{threshold_damping_1D}
    \tilde\nu^{\rm(cr,max)}\simeq2F^{\rm(cr,max)}\sqrt{\left(\kappa^{-1}+3\kappa^{-2}+3\kappa^{-3}\right){\rm e}^{-\kappa}},
\end{equation}
where the corresponding form factor is $F^{\rm(cr,max)}=1.28$. Equation (\ref{threshold_damping_1D}), which is plotted in
the inset of Fig.~\ref{fig7}b by the dotted line, also provides excellent agreement with the 2D model. Hence, this is a
convenient formula to identify conditions of absolute stability for a 2D crystal.

\section{Comparison with experiments}\label{conclusion}

\begin{figure*}
\includegraphics[width=0.9\textwidth,clip=]{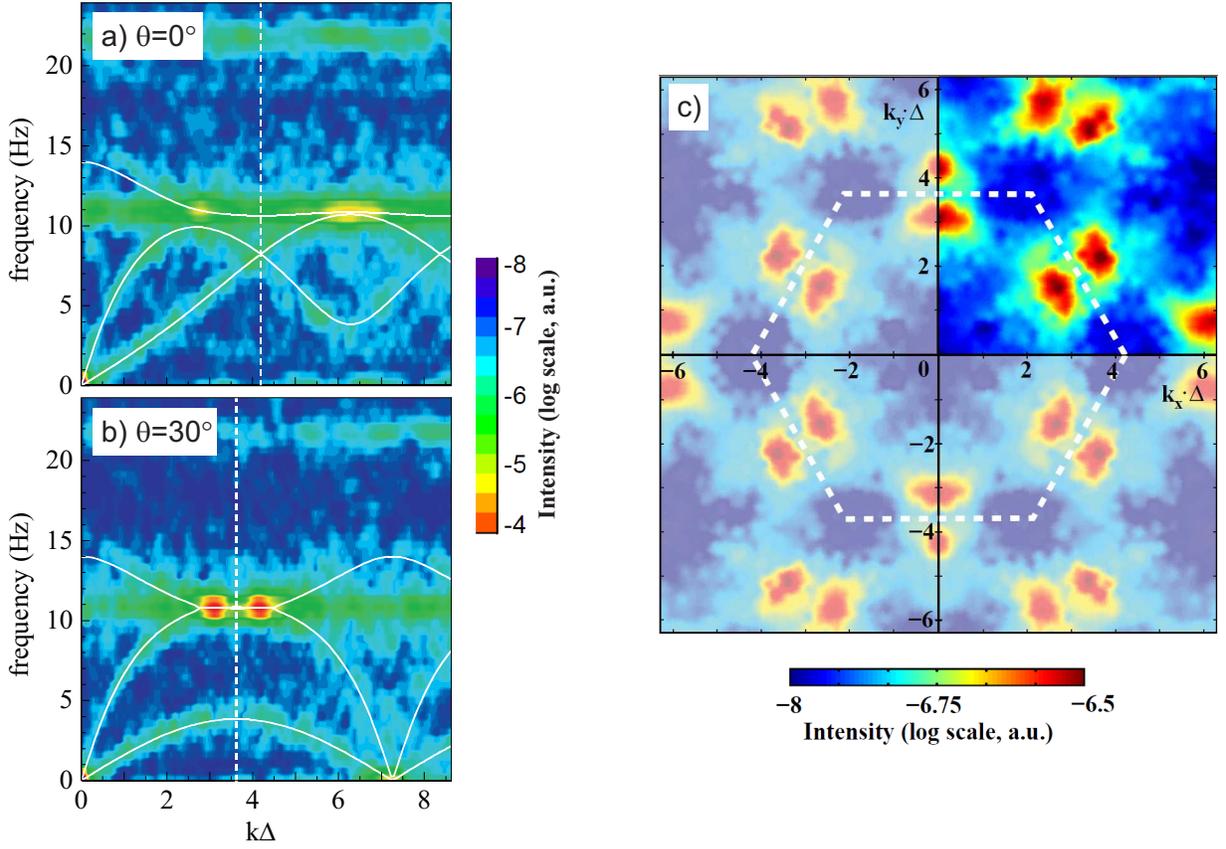}
\caption{In-plane fluctuation spectra for ``shallow'' mode intersection (experiment II \cite{PhysRevLett.104.195001} in
Table~\ref{tab1}). Shown are the DL modes (positive branches) for ${\bf k}$ at (a) $\theta=0^{\circ}$ and (b)
$\theta=30^{\circ}$, and (c) spectrum in the ${\bf k}$-plane integrated over frequency (in the range between 10.5~Hz~$\leq
f\leq$~14.5~Hz). The dashed lines show the border of the first Brillouin zones, solid lines in (a) and (b) are theoretical
curves for all tree principal DL modes (assuming $\tilde q=0.3$ and $\tilde\delta=0.33$). Fluctuation spectra in (a) and (b)
reveal clear traced of the out-of-plane mode (mixed polarization). In order to reduce noise in (c), the spectrum is an
average of six spectra corresponding to equivalent directions in the crystal (using the invariance of hexagonal lattice to
$60^{\circ}$ rotations). }\label{fig9}
\end{figure*}

Wake-induced mode coupling has been observed in several experiments
\cite{couedel:215001,PhysRevLett.104.195001,PhysRevLett.105.085004,PhysRevE.68.026405}. However, the analysis of
experimental results can be fastidious and needs to be done carefully. The fingerprints of the mode-coupling must be
identified properly in order to avoid misinterpretations of the results. Therefore, in this Section we use the experiment of
Ref.~\cite{PhysRevLett.104.195001} as a characteristic example to demonstrate the principal features of the wake-induced
mode coupling. Then, we give an overview of literature currently available on experiments with 2D crystals (stable and
unstable) \cite{couedel:215001,PhysRevLett.104.195001,PhysRevLett.105.085004,PhysRevE.68.026405,samsonov:026410}, the
summary of experimental parameters is given in Table~\ref{tab1}. Finally, we discuss the stability of crystals in terms of
thresholds introduced in the previous Section.

One of the most clear experimental evidence of mode-coupling in a 2D plasma crystal was reported in
Ref.~\cite{PhysRevLett.104.195001} (experiment II in Table~\ref{tab1}). A monolayer of microparticles was formed by
levitating them in a plasma sheath above the lower rf electrode in a capacitively coupled discharge in a (modified) GEC
chamber. The microspheres were made of melamine-formaldehyde and had a diameter of $9.19\pm0.09~\mu$m. The low argon
pressure of $0.76$~Pa ensured that the particle motion was only slightly damped. A high quality of the monolayer (with no
detected particles above or below) was verified by using a side-view camera. Details of the experimental procedure are given
in Refs.\cite{couedel:215001,PhysRevLett.104.195001}.

\begin{figure}
\includegraphics[width=7.5cm,clip=]{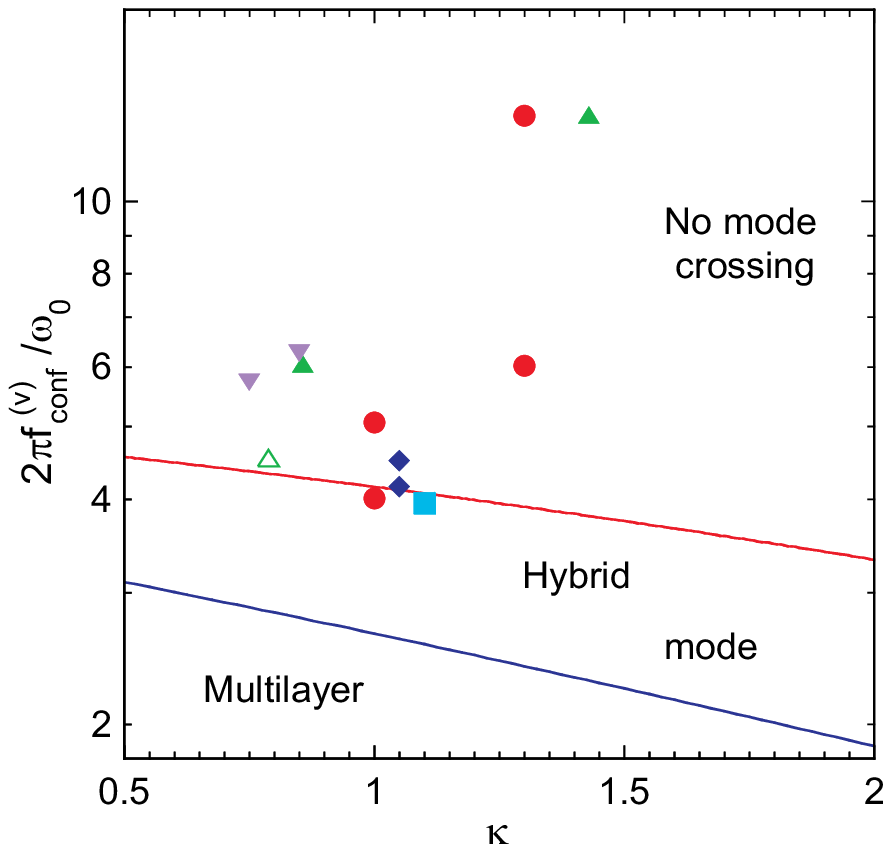}
\caption{Comparison of the vertical confinement frequency measured in experiments (Table~\ref{tab1}) with the theoretical
confinement threshold (Fig.~\ref{fig7}a): Blue diamonds are for experiments I and II, red circles for III-VI,
green up triangles for VII and VIII, open green up triangle for IX, blue square for X, and purple down triangles for XI and XII.
The vertical resonance frequency is normalized to $\omega_0=\sqrt{Q^2/m\Delta^3}$. For experiments I to VI, error bars are
available: $\pm15\%$ for $\kappa$, $\pm8\%$ for $f_{\rm conf}^{\rm(v)}$, and $\pm12\%$ for $\omega_0$.}
\label{fig10}
\end{figure}

The experimentally measured fluctuation spectra of two in-plane DL modes are shown in Fig.~\ref{fig9}. Since this experiment
was performed at conditions close to the onset of instability, all essential fingerprints of mode coupling are clearly
visible:
\begin{itemize}
\item{The mode coupling critically depends on the direction of the wave vector ${\bf k}$. The comparison of wave spectra
    calculated for $\theta=0^{\circ}$ (Fig.~\ref{fig9}a) and $\theta=30^{\circ}$ (Fig.~\ref{fig9}b) reveals that the
    onset of mode coupling first occurs at $\theta=30^{\circ}$ -- this results in the prominent ``hot spots'' in
    Fig.~\ref{fig9}b. Colored red, they are almost two orders of magnitude more intense than the ``normal'' branches of
    the fluctuation spectra. The hot spot that lies within the first Brillouin zone (left one) represents the hybrid
    mode, the other one is its image in the higher Brillouin zone. These observations are in agreement with theory and
    complimentary simulations (Fig.~\ref{fig3}), which predicts that the mode coupling first takes place at
    $\theta=30^{\circ}$ for ``shallow'' mode intersection.}
\item{The less intense spot seen in Fig.~\ref{fig9}a at $\theta=0^{\circ}$ and $k\Delta\simeq6.5$ is outside the first
    Brillouin zone (in Fig.~\ref{fig9}c the location is along $k_x$-axis, at the right edge). As explained in
    Sec.~\ref{remarks}, this spot should be mapped into the first Brillouin zone, where it represents the gap between
    two hot spots of the hybrid mode (in Fig.~\ref{fig9}c, along the border of the Brillouin zone at
    $\theta\simeq30^\circ$), and therefore its intensity is almost one order of magnitude smaller than the intensity of
    the hot spots.}
\item{Traces of mixed polarization (out-of-plane mode) are clearly seen next to the hybrid mode in Figs.~\ref{fig9}a and
    \ref{fig9}b. This gives a direct experimental proof that the out-of-plane mode becomes oblique close to the
    intersection point. A discrepancy of $\simeq5~\%$ between the experimentally measured in-plane projection of the
    out-of-plane mode and its theoretical dispersion relation can be explained by ``finite-$\tilde\delta$'' correction
    to the theory shown in Fig.~\ref{fig8} (which is about $4~\%$ for the chosen values of wake parameters).}
\item{Finally, all ``hot spots'' predicted by the theory and observed in simulations can be simultaneously seen in the
    experimental wave spectrum in ${\bf k}$-plane, which is shown in Fig.~\ref{fig9}c. }
\end{itemize}

Next, we analyze the stability of 2D plasma crystals in terms of the confinement threshold (which determines when the hybrid
mode emerges). Available literature data on stable and unstable crystals is summarized in Table~\ref{tab1}, and the measured
vertical confinement frequency $2\pi f^{\rm(v)}_{\rm conf}$ is compared to the theoretical threshold $\Omega ^{\rm
(cr)}_{\rm conf}$ in Fig.~\ref{fig10}. Note that for the comparison we change the normalization of $2\pi f^{\rm(v)}_{\rm
conf}$: Instead of using $\Omega_{\rm DL}$ -- a natural frequency scale in theory, now we employ the 2D analogue of dust
plasma frequency,
\begin{equation*}
    \omega_0=\sqrt{\frac{Q^2}{m\Delta^3}}\equiv\kappa^{-3/2}\Omega_{\rm DL}.
\end{equation*}
This normalization allows us to minimize the experimental errors, since the interparticle distance is usually measured with
much higher accuracy than the screening length (see footnote to Table~\ref{tab1}).

Figure~\ref{fig10} shows that all experiments listed in Table~\ref{tab1} as having no mode crossing are indeed represented
by points well above the confinement threshold line. To the contrary, the points for experiments marked by mode crossing are
located very close to the threshold line or below it. The only exception is the experiment VIII
\cite{PhysRevLett.105.085004} by Liu \textit{et al.}: The wake-induced mode coupling was clearly seen in the fluctuation
spectra measured in \cite{PhysRevLett.105.085004}, but the reported experimental parameters (yielding $2\pi f_{\rm
conf}^{\rm(v)}/\omega_0\simeq6$, marked by green up triangle in Fig.~\ref{fig10}) certainly do not allow the mode crossing,
which indicates large measurement errors (amended in the subsequent Erratum \cite{PhysRevLett.105.269901}). This example
demonstrates that, by comparing experimental parameters to the theoretical confinement threshold near the point where the
hybrid mode (hot spot) emerges in the fluctuation spectra one can perform very sensitive consistency test for the deduced
parameters.

\subsection{Effective charge and wake parameters}

At this point, let us briefly discuss the relation between the actual charge carried by particles and the effective charge
which determines the dispersion properties of DL waves in 2D plasma crystals.

The common method to determine $Q$ and $\lambda$ in 2D crystals is to fit the measured fluctuation spectra for the in-plane
modes by theoretical curves. The method is based on the {\it assumption} that the interparticle interaction potential can be
approximated by the Yukawa form \footnote{The functional form of the interaction potential can deviate from the Yukawa
dependence at large distances, but within several $\lambda$ this deviation is negligible indeed (see Ref.
\cite{kompaneets:052108}).} and provides rather high accuracy. However, the dispersion relations are determined by the
combined effect of direct interparticle interactions and the wake-mediated interactions. While the former contribution
scales as $\propto Q^2$, the latter depends on a particular form of the wake. For the model of a point-like wake charge used
in this paper the wake-mediated interactions scale as $\propto qQ$, which modifies the dispersion elements of the dynamical
matrix (\ref{elements1}). In the limit of small $\tilde\delta$ (which turns out to be a remarkably accurate approximation,
see Fig.~\ref{fig8}) this yields the common renormalizing factor $(1-\tilde q)$ in Eq. (\ref{elements1}), and therefore the
dispersion relations are determined by the {\it effective} charge $Q_{\rm eff}=\sqrt{1-\tilde q}\:Q$.

This fact should always be kept in mind when fitting the fluctuation spectra by theoretical dispersion relations: In theory,
frequencies are normalized by $\Omega_{\rm DL}\propto |Q|$ and hence the explicit dependence on the factor $(1-\tilde q)$
naturally disappears. This allows us, e.g., to directly compare the experimentally measured values of $2\pi f_{\rm
conf}^{\rm(v)}/\omega_0$ with the theoretical confinement threshold without knowing the wake parameters (see
Fig.~\ref{fig10}).

Of course, the comparison of theory and experiments in terms of the damping threshold requires knowledge of $\tilde q$ and
$\tilde\delta$, since the growth rate of the hybrid mode is directly proportional to the effective ``dipole moment'' of wake
$\tilde q\tilde\delta$. As was shown in Sec.~\ref{threshold2}, in order to compare the measured value of $\nu/\omega_0$
(last column of Table~\ref{tab1}) with the theoretical damping threshold plotted in Fig.~\ref{fig7}b, we need to known the
factor $\tilde q\tilde\delta/(1-\tilde q)$ [see Eq. (\ref{universal_norm}); additional factor $\sqrt{1-\tilde q}$ in the
denominator here is because $\omega_0$ in experiments is determined by $Q_{\rm eff}$]. In experiments, however, the wake
parameters are generally unknown, so that a reasonable assumption should always be made in order to perform the analysis in
terms of the damping threshold.

\subsection{Effect of ``finite'' $\tilde\delta$}
\label{large_delta}

As we mentioned in Sec.~\ref{threshold2}, under certain conditions the ``finite-$\tilde\delta$'' correction shown in
Fig.~\ref{fig8} can become important. In particular, this happens in ``marginal cases'', when the normalized value of the
confinement frequency is very close to the corresponding threshold line (viz., when the distance to the line is comparable
or less than the correction). For instance, such situation might occur for experiments II, VI, IX, and X shown in
Fig.~\ref{fig10}. However, typical experimental errors are relatively large, so that accounting for the corrections is
hardly required in these cases.

As regards the comparison of theory with numerical simulations, where all input parameters are known precisely, the
``finite-$\tilde\delta$'' corrections can become crucial. To illustrate this, let us consider simulations of the shallow
mode intersection shown in Fig.~\ref{fig3}. The (normalized) confinement frequency in this case is $\simeq1~\%$ {\it larger}
than $\widetilde\Omega_{\rm conf}^{\rm(cr)}$, i.e., according to the universal criterion for the confinement threshold the
system is still marginally stable. However, by taking into account the correction to the confinement threshold from
Fig.~\ref{fig8} (which is about $+4~\%$ in this case) we conclude that the system is in fact marginally unstable -- as
observed in the simulations. In contrast, for the case of the deep mode intersection shown in Fig.~\ref{fig5} (where the
confinement frequency is just $\simeq6.5~\%$ smaller) the correction plays practically no role.

\section{Conclusions and outlook}

The theory of mode-coupling instability \cite{PhysRevE.63.016409,zhdanov:zim} provides detailed picture of a {\it
plasma-specific} melting scenario operating in 2D plasma crystals. The melting associated with the wake-mediated coupling
between the longitudinal in-plane and out-of-plane modes can only be triggered if (i) the modes intersect and (ii) the
neutral gas damping is sufficiently low. Even if the instability is suppressed by the damping and the melting does not
occur, the coupling always results in the formation of a hybrid mode which is revealed in fluctuation spectra by anomalous
``hot spots'' emerging at distinct positions in the ${\bf k}$-plane. In the vicinity of the hybrid mode, one can observe
traces of mixed polarization for the two intersecting wave branches.

In this paper we showed that all these features can be considered as distinct fingerprints to identify the onset of the
wake-induced mode coupling. This, in turn, allows us to determine certain combinations of the crystal parameters with a
fairly good accuracy. For instance, the theory predicts a well-defined confinement threshold (for the ratio of the vertical
confinement frequency to the dust-lattice frequency) at which the ``hot spots'' emerge. It is noteworthy that the effect of
wakes on the dispersion relations is automatically taken into account via the charge renormalization, which yields the
effective charge -- the value which is deduced from the analysis of experimental fluctuation spectra.

On the other hand, the damping threshold as well as the evolution of the mode-coupling instability (if not suppressed) are
determined by parameters of the wake -- which are still poorly known for experiments. By identifying the onset of the
instability and comparing it with the damping threshold derived above, one could obtain effective ``dipole moment'' of the
wake. Furthermore, one could measure the energy growth rate at the initial instability stage and compare it with the
theoretical prediction. Systematic studies in this direction would be highly desirable. However, we note that such
measurements would only give us some mean characteristics of the wake, e.g., the first (``dipole'') moment in the multipole
expansion.

We believe that future research into the stability of 2D plasma crystals should be focused, in particular, on the
implementation of self-consistent wake models \cite{PhysRevE.64.046406,kompaneets:052108,Kompaneets08} for the calculation
of the mode coupling (which should also account for the spatial variation of the interaction parameters
\cite{Kompaneets05,Yaroshenko05}). Such studies will allow us to gain deeper insight into the plasma-specific mechanisms of
melting operating in 2D systems and provide more reliable basis for future research into generic melting mechanisms.

\begin{appendix}

\section{Elements of dynamical matrix $\textsf{D}$ [Eq. (\ref{matrix})]}\label{appendix1}

In Ref.~\cite{zhdanov:zim} we calculated the elements of $\textsf{D}$ with the accuracy $O(\tilde\delta)$. Although usually
$\tilde\delta$ is small indeed, such a linear expansion might not be always sufficient for some experimental conditions (see
Sec.~\ref{large_delta}). Below we present the results for arbitrary $\tilde\delta$. For brevity, we introduce the following
auxiliary functions:
\begin{equation}\nonumber
    \begin{array}{l}
        \Psi(x)=(x^{-1}+x^{-2}+x^{-3}){\rm e}^{-x},\\[.2cm]
        \Xi(x)=(x^{-1}+3x^{-2}+3x^{-3}){\rm e}^{-x},\\[.2cm]
        \Lambda(x)=\frac12[\Xi(x)-\Psi(x)].\\[.3cm]
    \end{array}
\end{equation}
Then the dispersion elements of Eq. (\ref{matrix}) are given by the following sums over integer $m$ and $n$ with excluded
$(0,0)$:
\begin{equation}\label{elements1}
    \begin{array}{l}
        \alpha_{\rm h}=\sum\limits_{m,n}\left\{\Psi(\kappa s)-\tilde q \left[\Psi(\kappa s_{\delta})-
            \tilde\delta^2s_{\delta}^{-2}\Xi(\kappa s_{\delta})\right]\right\}\\
            \hspace{5.8cm}\times\sin^2\frac12{\bf k}\cdot{\bf s},\\[.6cm]
        \alpha_{\rm v}=\sum\limits_{m,n}\left\{\Lambda(\kappa s)-\tilde q \left[\Lambda(\kappa s_{\delta})-
            \tilde\delta^2s_{\delta}^{-2}\Xi(\kappa s_{\delta})\right]\right\}\\
            \hspace{5.8cm}\times\sin^2\frac12{\bf k}\cdot{\bf s},\\[.6cm]
        \beta=\sum\limits_{m,n}\left[\Xi(\kappa s)-\tilde qs^2s_{\delta}^{-2}\Xi(\kappa s_{\delta})\right]\\
            \hspace{3.9cm}\times(s_y^2-s_x^2)s^{-2}\sin^2\frac12{\bf k}\cdot{\bf s},\\[.6cm]
        \gamma=\sum\limits_{m,n}\left[\Xi(\kappa s)-\tilde qs^2s_{\delta}^{-2}\Xi(\kappa s_{\delta})\right]
            s_xs_ys^{-2}\sin^2\frac12{\bf k}\cdot{\bf s}.
    \end{array}
\end{equation}
Here, the vector ${\bf s}$ with the components $s_x=\frac12m+n$ and $s_y=\frac{\sqrt{3}}2m$ and the absolute value
$s=\sqrt{m^2+mn+n^2}$ characterizes the relative positions of all neighbors in the hexagonal lattice; we also introduced
$s_{\delta}=\sqrt{s^2+\tilde\delta^2}$. The coupling elements are given by
\begin{equation}\label{elements2}
    {\textstyle\sigma_{x,y}=\tilde q\tilde\delta\sum\limits_{m,n}\Xi(\kappa s_{\delta})s_{x,y}s_{\delta}^{-2}\sin{\bf
    k}\cdot{\bf s}}.
\end{equation}
In the NN approximation (formally applicable for $\kappa\gg1$), the modes $\Omega_{\rm v}^2({\bf k})$ and $\Omega_{{\rm
h}\|}^2({\bf k})$ (whose coupling causes the hybridization) obtained from Eqs (\ref{elements1}) and (\ref{elements2})
coincide with Eqs. (\ref{Modes_1D})-(\ref{Coupling_coefficient_1D}) for a 1D string. In the limits $\tilde q\to0$ or
$\tilde\delta\to\infty$, Eqs (\ref{elements1}) and (\ref{elements2}) reduce to conventional (wake-free) results for DL modes
in 2D crystals. For $\tilde\delta\to0$ and finite $\tilde q$ one also obtains the wake-free results, but with the particle
charge renormalized by the factor $\sqrt{1-\tilde q}$.

\section{Hybrid mode for 1D model}\label{appendix1a}

The intersection point between the horizontal and vertical modes, $(\Omega_0,k_0)$, is readily derived from Eq.
(\ref{Modes_1D}):
\begin{equation*}
    \Omega_0=\sqrt{\frac{\kappa^2+2\kappa+2}{\kappa^2+3\kappa+3\kappa}}\;\Omega_{\rm conf},\quad
    \sin\frac12k_0=\frac{\Omega_{\rm conf}}{\Omega_{\rm conf}^{\rm(cr)}},
\end{equation*}
where $\Omega_{\rm conf}^{\rm(cr)}$ is given by the first Eq. (\ref{threshold_confinement_1D}) with $F_{\rm
conf}^{\rm(cr)}=1$. The maximum growth rate of the hybrid mode is obtained from Eq. (\ref{Coupling_1D}),
\begin{equation}\label{hybrid_1D}
    {\rm Im}~\Omega_{\rm hyb}(k_0)=\frac{\sigma(k_0)}{2\Omega_0},
\end{equation}
where $\sigma(k)$ is given by Eq. (\ref{Coupling_coefficient_1D}). This yields the critical damping rate
$\nu^{\rm(cr)}=\sigma(k_0)/\Omega_0$.

\section{Molecular dynamics simulations}\label{appendix2}

The charged particles were confined in a vertical parabolic well with an eigenfrequency which was varied in the range of
15~Hz~$<f_{\rm conf}^{\rm(v)} <$~25~Hz. In the horizontal plane, two options were investigated: parabolic confinement with
an eigenfrequency $f_{\rm conf}^{\rm(h)}<0.5$~Hz and periodic boundary conditions (with $f_{\rm conf}^{\rm(h)}=0$). The
equation of motion for each particle is
\begin{displaymath}
    m\ddot{\mathbf{r}}_i +m\nu \dot{\mathbf{r}}_i =\sum_{j\neq i}\mathbf{F}_{ij} + {\bf C}_{i} + \mathbf{L}_{i}
\end{displaymath}
where $m$ is the particle mass, $\nu$ is the damping rate, $\mathbf{r}_i$ the position of the $i$th particle, and ${\bf
C}({\bf r}_i)$ represents the confinement force. The force of interparticle interaction is
\begin{eqnarray}
    \mathbf{F}_{ij} & =  & -\frac{Q^2}{r_{ij}^2}\exp\Big({-\frac{r_{ij}}{\lambda}}\Big)\Big(1+\frac{r_{ij}}{\lambda}\Big)
    \frac{\mathbf{r}_{ij}}{r_{ij}}\nonumber \\
                    &    & +\frac{Qq}{r_{{\rm w}_{ij}}^2} \exp\Big({-\frac{r_{{\rm w}_{ij}}}{\lambda}}\Big)\Big(1+
    \frac{r_{{\rm w}_{ij}}}{\lambda}\Big)\frac{\mathbf{r}_{{\rm w}_{ij}}}{r_{{\rm w}_{ij}}}\nonumber
\end{eqnarray}
where ${\bf r}_{ij}$ is the distance from the particle $i$ to the particle $j$, ${\bf r}_{{\rm w}_{ij}}$ is the distance to
the wake of the particle $j$, $Q$ is the particle charge, and $q$ is the point-like charge of the wake placed at a distance
$\delta$ below each particle. The Langevin force is defined as
\begin{displaymath}
    \langle\mathbf{L}_i(t)\rangle={\bf 0},\qquad \langle\mathbf{L}_i(t+\tau)\mathbf{L}_j(t)\rangle = 2\nu
    mT\delta_{ij}\delta(\tau),
\end{displaymath}
where $T$ is the temperature of the thermostat (here, $\delta$ is the Kronecker delta or the Dirac delta function).

In order to simulate systems of a size comparable to experimental (with $N\gtrsim10^4$ particles) without any approximation
(i.e., without cutoff radius in the interactions), the code was written in CUDA C and run on NVIDIA Tesla C1060 GPU or C2050
GPU computing cards. The equations of motion were solved using Beaman algorithm with predictor-corrector.

\section{Calculation of thresholds for 2D model}\label{appendix3}

To derive the confinement threshold, we substitute the dispersion elements from Eq. (\ref{elements1}) in Eq.
(\ref{threshold_confinement1}) which yields
\begin{widetext}
\begin{equation}
    \Omega_{\rm conf}^{\rm(cr)}=\sqrt{\sum\limits_{m,n}\eta_m\left\{\left(1+p\right)
    \left[\Xi(\kappa s)-\tilde q\Xi(\kappa s_{\delta})\right]+\tilde q\tilde\delta^2s_{\delta}^{-2}\left(3+p\right)
    \Xi(\kappa s_{\delta})\right\}}.\label{app_confinement1}
\end{equation}
\end{widetext}
Here the fact that $|{\bf k}_{\rm cr}|s_y=m\pi$ is taken into account, so that it is convenient to introduce the parameter
$\eta_m=\frac12 \left[1-(-1)^m\right]$ (which is equal to 1 or 0 for odd or even $m$, respectively). Furthermore, for
brevity we use $p\equiv s^{-2}(s_x^2-s_y^2)$.

For the accuracy $O(\tilde\delta)$, we can neglect in Eq. (\ref{app_confinement1}) the second term in curly braces and set
$s_{\delta}=s$, which results in the following expression:
\begin{equation}
    \frac{\Omega_{\rm conf}^{\rm(cr)}}{\sqrt{1-\tilde q}}=\sqrt{\sum\limits_{m,n}\eta_m\left(1+p\right)\Xi(\kappa s)},
    \label{app_confinement2}
\end{equation}
where the r.h.s. is a function of $\kappa$ only. Thus, Eq. (\ref{app_confinement2}) determine the universal dependence of
$\Omega_{\rm conf}^{\rm(cr)}$ on $\kappa$.

The minimum value of the confinement frequency defined in Eq. (\ref{threshold_confinement3}) is
\begin{widetext}
\begin{equation}
    \Omega_{\rm conf}^{\rm(min)}=\sqrt{2\sum\limits_{m,n}\eta_m\left\{\Lambda(\kappa s)-\tilde q\Lambda(\kappa s_{\delta})
    +\tilde q\tilde\delta^2s_{\delta}^{-2}\Xi(\kappa s_{\delta})\right\}},\label{app_confinement3}
\end{equation}
\end{widetext}
for the accuracy $O(\tilde\delta)$ it is reduced to
\begin{equation}
    \frac{\Omega_{\rm conf}^{\rm(min)}}{\sqrt{1-\tilde q}}=\sqrt{2\sum\limits_{m,n}\eta_m\Lambda(\kappa s)}.\label{app_confinement4}
\end{equation}

For the damping threshold, we substitute the coupling element [Eq. (\ref{elements2})] in Eq. (\ref{threshold_pressure1}) and
the dispersion elements in Eqs  (\ref{threshold_pressure1a}) and (\ref{threshold_pressure2}). After some algebra we get
\begin{widetext}
\begin{eqnarray}
    \nu^{\rm(cr)} & = &
        \frac{\tilde q\tilde\delta\sum\limits_{m,n}s_{\delta}^{-2}s_y\Xi(\kappa s_{\delta})\sin|{\bf k}_0|s_y}{\sqrt{\sum\limits_{m,n}
        \left\{\Psi(\kappa s)-\tilde q \Psi(\kappa s_{\delta})+p\left[\Xi(\kappa s)-\tilde q\Xi(\kappa s_{\delta})\right]
        +\tilde q\tilde\delta^2s_{\delta}^{-2}\left(1+p\right)\Xi(\kappa s_{\delta})\right\}\sin^2\frac12|{\bf k}_0|s_y }}
        \label{app_pressure1}, \\
    \Pi & = &
        \sum\limits_{m,n}\left(\eta_m-\sin^2{\textstyle\frac12}|{\bf k}_0|s_y\right)\left\{\left(1+p\right)\left[\Xi(\kappa s)
        -\tilde q\Xi(\kappa s_{\delta})\right]+\tilde q\tilde\delta^2s_{\delta}^{-2}\left(3+p\right)
        \Xi(\kappa s_{\delta})\right\}. \label{app_pressure2}
\end{eqnarray}
\end{widetext}
For the accuracy $O(\tilde\delta)$, we neglect the last terms in curly braces and set $s_{\delta}=s$. Then Eqs
(\ref{app_pressure1}) and (\ref{app_pressure2}) are reduced to
\begin{widetext}
\begin{eqnarray}
    \frac{\sqrt{1-\tilde q}}{\tilde q\tilde\delta}\nu^{\rm(cr)} & = &
        \frac{\sum\limits_{m,n}s^{-2}s_y\Xi(\kappa s)\sin|{\bf k}_0|s_y}{\sqrt{\sum\limits_{m,n}\left[\Psi(\kappa s)+p\Xi(\kappa s)\right]
        \sin^2\frac12|{\bf k}_0|s_y }},\label{app_pressure3}\\
    \frac{\Pi}{1-\tilde q} & = &
        \sum\limits_{m,n}\left(\eta_m-\sin^2{\textstyle\frac12}|{\bf k}_0|s_y\right)\left(1+p\right)\Xi(\kappa s),\label{app_pressure4}
\end{eqnarray}
\end{widetext}
where the r.h.s. depend only on $\kappa$ and $|{\bf k}_0|$. Thus, Eqs (\ref{app_pressure3}) and (\ref{app_pressure4})
determine the universal dependence of the critical damping rate $\nu^{\rm(cr)}$ on the confinement control parameter $\Pi$
via the parametric dependence on $|{\bf k}_0|$.

\end{appendix}


%

\end{document}